\documentclass[twocolumn]{aastex6}

\newcommand{\rrab}{RR{\sl ab}}
\newcommand{\rrc}{RR{\sl c}}

\shorttitle{RR Lyrae Stars in DECam bands}
\shortauthors{Vivas et al.}

\begin{document}

\title{Absolute Magnitudes and Colors of RR Lyrae stars in DECam Passbands from Photometry of the Globular Cluster M5}

\author{A. Katherina Vivas\altaffilmark{1},  Abhijit Saha\altaffilmark{2}, Knut Olsen\altaffilmark{2}, Robert Blum\altaffilmark{2},
Edward W. Olszewski\altaffilmark{3},  Jennifer Claver\altaffilmark{2},  Francisco Valdes\altaffilmark{2}, Tim Axelrod\altaffilmark{3},  
Catherine Kaleida\altaffilmark{5}, Andrea Kunder\altaffilmark{4}, Gautham Narayan\altaffilmark{2}, Thomas Matheson\altaffilmark{2},  
Alistair Walker\altaffilmark{1} }

\altaffiltext{1}{Cerro Tololo Inter-American Observatory, National Optical Astronomy Observatory, Casilla 603, La Serena,
Chile, \email{kvivas@ctio.noao.edu}}
\altaffiltext{2}{National Optical Astronomy Observatory, 950 North Cherry Avenue, Tucson, AZ, 85719}
\altaffiltext{3}{University of Arizona, Steward Observatory, 933 North Cherry Avenue, Tucson, 85721}
\altaffiltext{4}{Leibniz-Institut f\"ur Astrophysik Potsdam (AIP), An der Sternwarte 16, D-14482 Potsdam, Germany}
\altaffiltext{5}{Space Telescope Science Institute, 3700 San Martin Drive, Baltimore, MD 21218}
\email{kvivas@ctio.noao.edu}

\begin{abstract}
We characterize the absolute magnitudes and colors of RR Lyrae stars in the globular cluster M5 in the $ugriz$ filter system of the Dark Energy
Camera (DECam). We provide empirical Period-Luminosity (P-L) relationships in all 5 bands based on 47 RR Lyrae stars of the type ab and 14 stars of the
type c. The P-L relationships were found to be better constrained for the fundamental mode RR Lyrae stars in the $riz$ passbands, with dispersion of
0.03, 0.02 and 0.02 magnitudes, respectively. The dispersion of the color at minimum light was found to be small, supporting the use of this 
parameter as a means to obtain accurate interstellar extinctions along the line of sight up to the distance of the RR Lyrae star.  We found a trend
of color at minimum light with pulsational period that, if taken into account, brings the dispersion in color at minimum light to $\leq 0.016$ magnitudes
for the $(r-i)$, $(i-z)$, and $(r-z)$ colors. These calibrations will be very useful for using RR Lyrae stars from DECam observations as 
both standard candles for distance determinations and color standards for reddening measurements.
\end{abstract}


\section{Introduction}\label{sec-intro}

RR Lyrae stars are well established standard candles and tracers of
old stellar populations throughout the Local Group. They are also
extremely good standard colors.  As first demonstrated by
\citet{Sturch66}, the colors of fundamental mode RR Lyrae stars near the minimum light phase of their pulsation cycle are predictable with accuracies of a few percent.  As such, they broadcast both their distance as well as their line of sight reddening, and hence extinction. This property makes RR Lyrae stars specially good probes in regions of high interstellar extinction such as the Galactic center and inner Bulge, 
where they exist in large numbers.  There is a rich history of
attempts to study aspects of the inner Galaxy using RR~Lyrae stars
\citep[][among
others]{baade46,plaut68,plaut70,plaut73,blanco84,blanco92,alcock98,kunder08,gran16}. The
most recent and extensive photometric catalogs have come from the OGLE
project \citep[e.g.,][]{Udalski15}.  Their RR Lyrae survey and results are described in \citet{Pietr15}.  Relative to these past investigations, the Dark Energy Camera \citep[DECam,][]{flaugher15} on the Blanco 4m telescope offers a more extensive chromatic coverage, allowing us to measure colors over a wide spectral range. This enables us to not only measure reddening to the RR Lyrae stars accurately, but also to examine alleged variations in the reddening law towards the Galactic center.  In addition to simply acquiring a more accurate map of the RR Lyrae stars' spatial distribution (and by extension that of the `ancient' stellar population), the relatively high density of RR~Lyrae stars in these regions offers a chance to make an accurate map of the line of sight extinction through the foreground disk that depends only upon our knowledge of RR Lyrae star colors. This should be an improvement over prior methods such as using colors of red clump stars,  which are much more sensitive to stellar population parameters such as metallicity.  If the goal is to examine the stellar populations in the bulge and bar using color-magnitude and Hess diagrams, it is desirable to correct for reddening using a probe like RR Lyrae star minimum light colors, which as we discuss later, is known to be insensitive to metallicity. 

To exploit this opportunity, we must first determine empirically the intrinsic colors of the RR~Lyrae stars at minimum light in the photometric system of the DECam passbands.
As the globular cluster M5 (NGC 5904) is a well studied cluster rich
in RR Lyrae stars \citep[][and references therein]{clement01} and is located towards the Southern Galactic cap with minimal line of sight reddening, its RR Lyrae stars are ideal for establishing baseline colors and absolute magnitudes that can be used to interpret photometry of reddened RR Lyrae stars towards the Galactic center and along or near the plane of the Galaxy. The characteristics of M5, including distance and reddening, are summarized in \citet{harris96}\footnote{Updated information is maintained on a web-page: \url{http://physwww.physics.mcmaster.ca/~harris/mwgc.dat}}.
 In this paper, we present results from DECam photometry of M5 RR Lyrae stars to establish their colors at minimum light and absolute magnitudes in the DECam passbands, which are 
 needed to determine the extinction toward the stars and their distances.
 In upcoming papers we will use these results as established base-line properties of RR Lyrae stars to study the Galactic Bulge. 
 We note that \citet{ngeow17} recently presented an analysis of the
 Sloan Digital Sky Survey (SDSS) colors of RR Lyrae stars at maximum and minimum light 
based on Stripe 82 data. The advantage of our study is that since we have RR Lyrae stars residing in the same cluster, we are able to probe absolute magnitudes and colors at a fixed distance, greatly minimizing distance uncertainties from the equation.

The paper is laid out as follows.  The imaging data, processing  and photometry are described in \S\ref{sec-data}. A description of the periodic
variable stars identified in this work, namely 66 RR Lyrae stars and 1 SX Phe star, are shown in section \S\ref{sec-RRLS}, and the location of these stars
in the Color-Magnitude diagram (CMD) of M5 is discussed in Section \S\ref{sec-CMD}.  Section \S\ref{sec-color} 
presents the average colors at minimum light of both the fundamental
mode (\rrab) and first overtone (\rrc) RR Lyrae stars in different
filters, as well as discussing dependence with period and metallicity.
Section \S\ref{sec-M} provides Period-Luminosity (P-L) relationships for the RR Lyrae stars in M5 in the DECam $ugriz$ system. Conclusions are provided in \S\ref{sec-conclu}.

\section{Data \& Photometry}\label{sec-data}

\begin{figure}[htb!]
\plotone{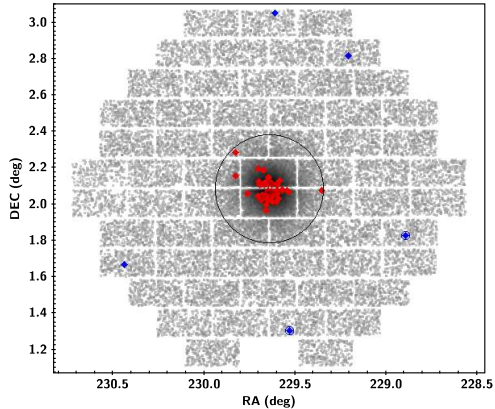}
\caption{Density map of the 60,335 objects detected in the field of view of DECam centered on M5 in the $gri$ filters.
For reference, the black circle marks $10 \, r_h$ of the cluster, with $r_h=1\farcm 77$ \citep{harris96}.
Colored diamonds show the location of the periodic variable stars identified in this work. All the red diamonds matched with known stars 
in the catalog of variable stars of M5 \citep{clement01}. The blue diamonds did not have an entry in that catalog. Because of their 
distance from the center of the cluster and their faint magnitudes, they should be distant halo stars. 
The diamonds encircled in blue are two new discoveries.}
\label{fig-M5sky}
\end{figure}

Observations were obtained during 2013 and 2014 with the DECam imager on the 4m Blanco Telescope at Cerro Tololo Inter-American 
Observatory (CTIO), Chile.
Repeated DECam images of a field centered on M5 (RA=15:18:33.2, DEC=+02:04:51.7, J2000.0) were obtained using the $u,g,r,i$, and $z$ filters. 
The large field of view (FOV) of DECam ($2\fdg 2$) easily covers the whole globular cluster with only the central CCDs of the camera (Figure~\ref{fig-M5sky}).  
The journal of observations (date (evening) and number of observations per filter) are recorded in Table~\ref{tab-observations}. M5 is a nearby globular cluster located at only 7.5 kpc from the Sun,
whose RR Lyrae stars are henceforth expected to be bright, at V magnitudes of $\sim 15.1$ \citep{harris96}. 
Exposure times were therefore short in our images with 30s in the $u$ band and 
just 10s in all the other filters. On photometric nights,
observations were also made of photometric standards recently established by \citet{Narayan16}.
For RR Lyrae stars, it is best when the targets can be observed several 
times throughout the night, because periods are less than a day.  The
observations of M5 were made, however, on assigned nights when fields near the Galactic center 
were the primary targets. As a result of the large difference in right ascension between M5 and the Galactic bulge fields, M5 could only be observed for part of the night.  In
addition, many of the observations reported in Table~\ref{tab-observations} were taken continuously with separations of only minutes.
As periods of RR Lyrae stars range from   
0.2 to 0.9 days, that small separation in time does not enable proper
sampling of the light curves. This non-optimal cadence 
has some adverse effects for sampling the light curves, which we discuss later in the paper. Nevertheless, the observations allow for light curves that trace the pulsation of the star well (see Figure~\ref{fig-lightcurve}).

\begin{deluxetable}{lccccc}
\tablecolumns{6}
\tablewidth{0pc}
\tablecaption{Number of DECam observations of M5 \label{tab-observations}}
\tablehead{
Date  &  $N_u$  &  $N_g$   & $N_r$  & $N_i$  & $N_z$ \\
}
\startdata
2013 Jun 7 & 2 & 2 & 3 & 2 & 2 \\
2013 Jun 8 & 8 & 8 & 13 & 8 & 8 \\
2013 Jun 9 & 11 & 12 & 17 & 13 & 14 \\
2013 Jun 21 & 1 & 3 & 6 & 3 & 4 \\
2014 Mar 7 & 8 & 8 & 8 & 8 & 8 \\
2014 Mar 8 & 6 & 6 & 6 & 6 & 6 \\
2014 Mar 9 & 12 & 12 & 12 & 12 & 12 \\
\hline
TOTAL & 50 & 53 & 68 & 54 & 56 \\
\enddata
\end{deluxetable}

The data were processed through the Community Pipeline 
\citep{Valdes14} for bias subtraction, flat-fielding, bad pixel mask generation,  and WCS refinement. Reduced images are publicly
available through the NOAO Science Archive\footnote{\url{https://archive.noao.edu}}.
Point-Spread function (PSF) fitting photometry was performed on the reduced images using a variant of DoPHOT \citep{Schechter93}.  DoPHOT was embedded inside an integrated IDL driven 
procedure described in \S3.2 of \citet{Saha10} for the MOSAIC imager, so that the final product is a list of objects with aperture corrected  instrumental 
magnitudes. The only differences in the procedure for DECam are that:
\begin{enumerate}
\item Instead of laying the CCD by CCD sections onto a common gnomonic projection, creating one single image for the whole field, the photometry was performed 
separately on each CCD of the DECam images, keeping memory utilization manageable, and 
\item Aperture corrections were calculated independently for each
  CCD. Unlike for MOSAIC, as a result of the excellent optico-mechanical layout of DECam, and the use of 
the hexapod for focus and alignment,  no variation in aperture corrections as a function of position in the field can be seen within the area covered by an individual 
CCD. As the aperture corrections are calculated independently for each CCD, any PSF variations over large changes in field position are effectively 
accommodated on a CCD-to-CCD granularity.
\end{enumerate}

The instrumental magnitudes were calibrated to the system defined by \citet{Narayan16} using observations of two of the standard stars from that paper that were observed on photometric nights contemporaneously with the observations presented here.  As described in that paper, these magnitudes are defined using spectrophotometric fluxes for the standards, convolved through DECam passbands and typical atmospheric extinction at CTIO, and are therefore native to the DECam instrument, analogous, but {\emph not} identical to the SDSS system.

Absolute photometry was checked by comparing with the SDSS photometry of M5 provided by \citet{an08}. For this comparison we transformed our catalog to the 
SDSS system using equations derived by the Dark Energy Survey Collaboration (D. Tucker, private communication). The offsets between the two sets are 0.05, 0.02,
0.003, 0.03 and 0.02 mag in $ugriz$, respectively.

A total of $16.2 \times 10^6$ individual measurements were processed and stored in a MySQL database. The photometry was 
cleaned by eliminating measurements if: (1) they were within 50 pixels from the borders of the CCDs,
(2) there were two or more cosmic rays detected near the object within a radius of the FWHM of the major axis of the fitted PSF,  (3) other objects were found within a radius of the FWHM of the major axis of the fitted PSF the object and their contribution relative to the object's 
flux is  $>0.3$ magnitudes, (4) the fitted sky value for an object is 
$<$ SKYLOW or $> 3\times$SKYHI, where SKYLOW and SKYHI are the 2nd and 90th percentile, respectively, of the distribution of fitted sky values of all objects in the CCD in which the 
object is found, (5) the DoPHOT reported photometric error is higher than twice  the average reported error of all stars for that image that share the same half magnitude bin,
and (6) the photometric error of a measurement is significantly higher that the average of the errors of all measurements for the same star. Specifically, 
we eliminated measurements if the error was $>\langle \rm error \rangle + 3.5\, \sigma_{\rm error}$, where $\langle \rm error \rangle$ is the 
average  of the photometric error of all measurements for that star and $\sigma_{\rm error}$ is the standard deviation of the error distribution. 
Cosmic rays are identified, and affect pixels masked, by the DoPHOT program as part of the photometry measuring process. 
The elimination of stars with more than 2 cosmic rays in its immediate neighborhood from further analysis affects less than 2.4\% of the detected stars.
We also required that there were at least 5 good measurements per object in each filter. 
The final number of objects in our catalogs ranged from 43,792 in the
$u$ band to 176,704 objects in the $r$ band.

\section{RR Lyrae stars in M5}\label{sec-RRLS}

\begin{figure}
\includegraphics[width=0.5\textwidth]{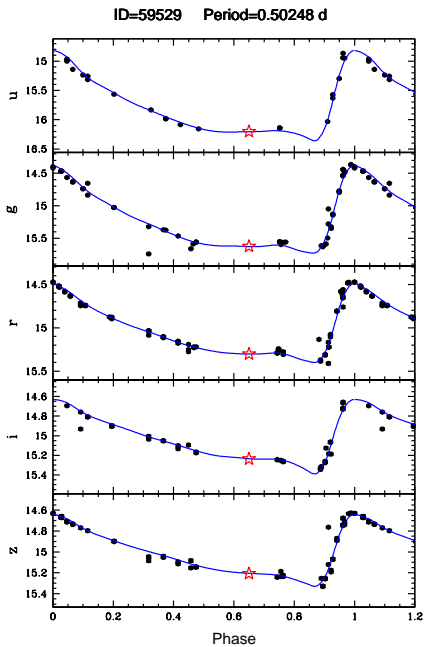}
\caption{Phased light curve in 5 bands for star 59529. The red star shows the magnitude at minimum light ($m_{\rm ref}$)
which, for {\rrab } stars like the one above, was measured at phase $\phi = 0.65$.  The blue line is the template, from the library of \citet{sesar10}, that best fitted
the data in each band. 
The complete figure set for all the periodic variable stars (67 figures) is available in the online journal.}
\label{fig-lightcurve}
\end{figure}

RR Lyrae stars were identified in the data by using the procedure described in detail in Saha 2017 (in preparation). 
Here, we summarize the most important details. Variable stars were flagged using a bootstrap $\chi^2$ using 
DoPHOT reported photometric errors. A visual tool that blinks between the brightest and faintest image for each star
enabled confirmation or rejections of the variability flag assignment. More than 90\% of the flagged cases were visually confirmed.
Then, periodic stars were searched using the Psearch algorithm \edit1{\citep{saha17}} that combines Fourier transforms and the 
classical Lafler \& Kinman method \citep[a string-length method,][]{Lafler65} in the 5 photometric bands simultaneously, 
which results in a powerful tool for sparse data. The resulting periodogram for each star was visually inspected interactively,
allowing exploration of the different peaks in the diagram. Periodic variable stars were then selected based on 
the shape of their light curve, period, and amplitude. A total of 66 RR Lyrae stars and 1 SX Phe were recognized in the field of M5.
The individual measurements for the periodic variable stars are provided in Table~\ref{tab-lcdata} as an electronic table. The table contains ID, MJD, filter, magnitude, and its error.
Lightcurves for all the stars are provided as online only material. Figure~\ref{fig-lightcurve} shows an example for one of the stars.

\begin{deluxetable}{lcccc}
\tablecolumns{5}
\tablewidth{0pc}
\tablecaption{Individual measurements in $ugriz$ of the periodic variables in the field of M5 \label{tab-lcdata}}
\tablehead{
ID & MJD & Filter & Mag & Error \\
}
\startdata
   16339 &  56451.722332 & u & 18.552 &  0.027 \\
   16339 &  56451.723139 & u & 18.590 &  0.024 \\
   16339 &  56452.499995 & u & 18.703 &  0.028 \\
   16339 &  56452.500658 & u & 18.661 & 0.030 \\
   16339 &  56452.613786 & u & 18.287 &  0.022 \\
\enddata
\tablecomments{Table~\ref{tab-lcdata} is published in its entirety in the electronic edition of The 
Astronomical Journal. A portion is shown here for guidance regarding its form and content.}
\end{deluxetable}  

The non-optimal cadence of our observations prevented determination of
the correct period for some of the RR Lyrae
stars. We took advantage of the fact that the variable stars in M5 have been extensively studied before and adopted
in several cases the published periods in the catalog of \citet{clement01}, which in the case of M5 had an 
update\footnote{\url{http://www.astro.utoronto.ca/~cclement/read.html}} in June 2015. A recent revision of variable stars in M5
is also provided by \citet{arellano16}.
Each light curve was fitted with a template from the library of
\cite{sesar10} that was created from SDSS $ugriz$ observations of RR Lyrae stars in Stripe 82. To find the optimal fit, we allowed small variations on the period and on the observed amplitude, maximum magnitude, and initial phase, in the range of $\pm 0.001$ d, $\pm 0.2$ mag, $\pm 0.2$ mag, and 0.2 units of phase, respectively. These fitted templates turned out to be very useful to determine the magnitude at minimum phase, as explained 
in \S~\ref{sec-color}. 

\floattable
\begin{deluxetable}{lccccccccccccccl}
\tabletypesize{\small}
\rotate
\tablecolumns{16}
\tablewidth{0pc}
\tablecaption{Periodic variable stars identified in the M5 field \label{tab-RRLS}}
\tablehead{
ID & RA & DEC & Period & Type & $\langle u \rangle$ & $\langle g \rangle$ & $\langle r \rangle$ & $\langle i \rangle$ & $\langle z \rangle$ &
$u_{\rm ref}$ & $g_{\rm ref}$ & $r_{\rm ref}$ & $i_{\rm ref}$ & $z_{\rm ref}$ & Other Identification \\
 & (J2000.0) & (J2000.0) & (d) & & & & & & & & & & & & \\
}
\startdata
113495 & 15:15:34.11 & +01:49:30.5 & 0.63013 &     ab & 18.43 &  17.82 &  17.57 &  17.51 &  17.46 &  18.66 &  18.04 &  17.69 &  17.61 &  17.56 & NEW        \\ 
114707 & 15:16:49.73 & +02:48:42.9 & 0.65687 &     ab & 18.72 &  18.09 &  17.84 &  17.78 &  17.74 &  19.03 &  18.33 &  18.01 &  17.90 &  17.86 & CRTS\_J151649.7+024841 \\ 
 72475 & 15:17:24.23 & +02:04:24.1 & 0.34908 &      c & 15.19 &  14.74 &  14.65 &  14.69 &  14.67 &  15.58 &  15.11 &  14.88 &  14.89 &  14.83 & V67                    \\ 
139277 & 15:18:06.77 & +01:18:10.5 & 0.53037 &     ab & 19.93 &  19.35 &  19.21 &  19.21 &  19.16 &  20.51 &  19.73 &  19.45 &  19.39 &  19.35 & NEW                    \\ 
 74636 & 15:18:08.04 & +02:03:45.8 & 0.45141 &     ab & 15.73 &  15.12 &  15.05 &  15.11 &  15.09 &  16.11 &  15.54 &  15.33 &  15.32 &  15.27 & V29                    \\ 
\enddata
\tablecomments{Table~\ref{tab-RRLS} is published in its entirety in the electronic edition of The 
Astronomical Journal. A portion is shown here for guidance regarding its form and content.}
\end{deluxetable}  

Out of the 129 RR Lyrae stars registered in
M5 in the catalog of \citet{clement01}, we recovered 62 (red diamonds in Figure~\ref{fig-M5sky}). The remaining 68 stars that we did not recover were located 
mostly toward the center of M5, which lies in a gap among the CCDs in our DECam images. There are 5 stars in our catalog that were not present in the M5 catalog of 
variable stars of \citet{clement01} that are shown as blue diamonds in Figure~\ref{fig-M5sky}. 
None of them is expected to be a member of 
the cluster because they are outside a radius of $10 \, r_h$ and they have mean $g$ magnitudes between 17.8 and 19.3, 
which locate them much farther away than M5. Thus, 
most likely they are distant halo stars lying along the line of sight. Three of those halo stars appear in the Catalina Real-Time Transient Survey 
\citep[CRTS][]{drake13a,drake13b,drake14} and were classified as RR Lyrae stars, but the other two, namely 113495 and 139277, have not been cataloged before and should be considered new discoveries.

In Table~\ref{tab-RRLS} we present the list of periodic variable stars. It contains ID, Right Ascension (RA), Declination (DEC), period in 
days, type of pulsating star, mean magnitudes in $ugriz$, and  magnitudes at minimum light ($m_{\rm ref}$)  in $ugriz$ that were
measured as explained in section \S\ref{sec-color}.
The last column provides the cross-identification with the V* identification in \citet{clement01} or the ID in the CRTS. The two new stars
are denoted as ``NEW.''
The reported mean magnitudes in Table~\ref{tab-RRLS} are not plain averages but were obtained by integrating the best fitted template of the light curve in intensity units, and transforming back to magnitudes.
This method of calculating the mean magnitudes avoids the biases toward 
minimum magnitudes (because the RR Lyrae stars spend most of their pulsation cycle time at minimum light) as well as
those biases that may appear as a result of unevenly sampled lightcurves. 

We estimated the photometric errors in each band by determining the standard deviation of all of the individual magnitudes for each star in each CCD. 
As most of the stars in the field are non-variable, the standard deviation provides a good estimate of the real photometric errors. 
In the CCDs containing the bulk of the M5 stars, which are representative of the crowded environment of our observations, these errors
average 0.013, 0.011, 0.011, 0.012 and 0.012 magnitudes at the level of the horizontal branch (HB) in the $u$, $g$, $r$, $i$ and $z$ bands, respectively.

\section{Color-Magnitude Diagrams} \label{sec-CMD}

\begin{figure*}[ht]
\centering
\includegraphics[width=0.49\textwidth]{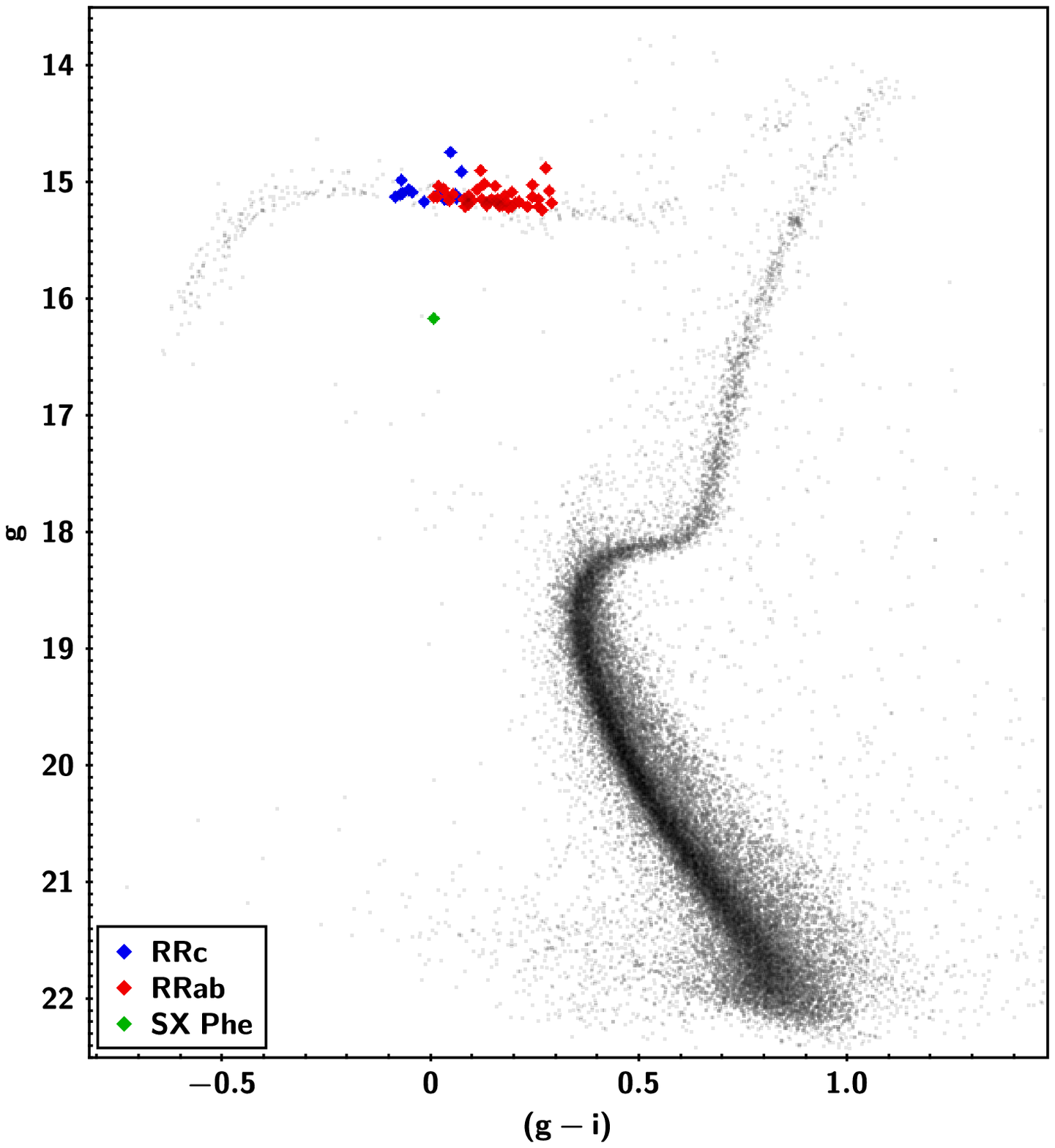}
\includegraphics[width=0.49\textwidth]{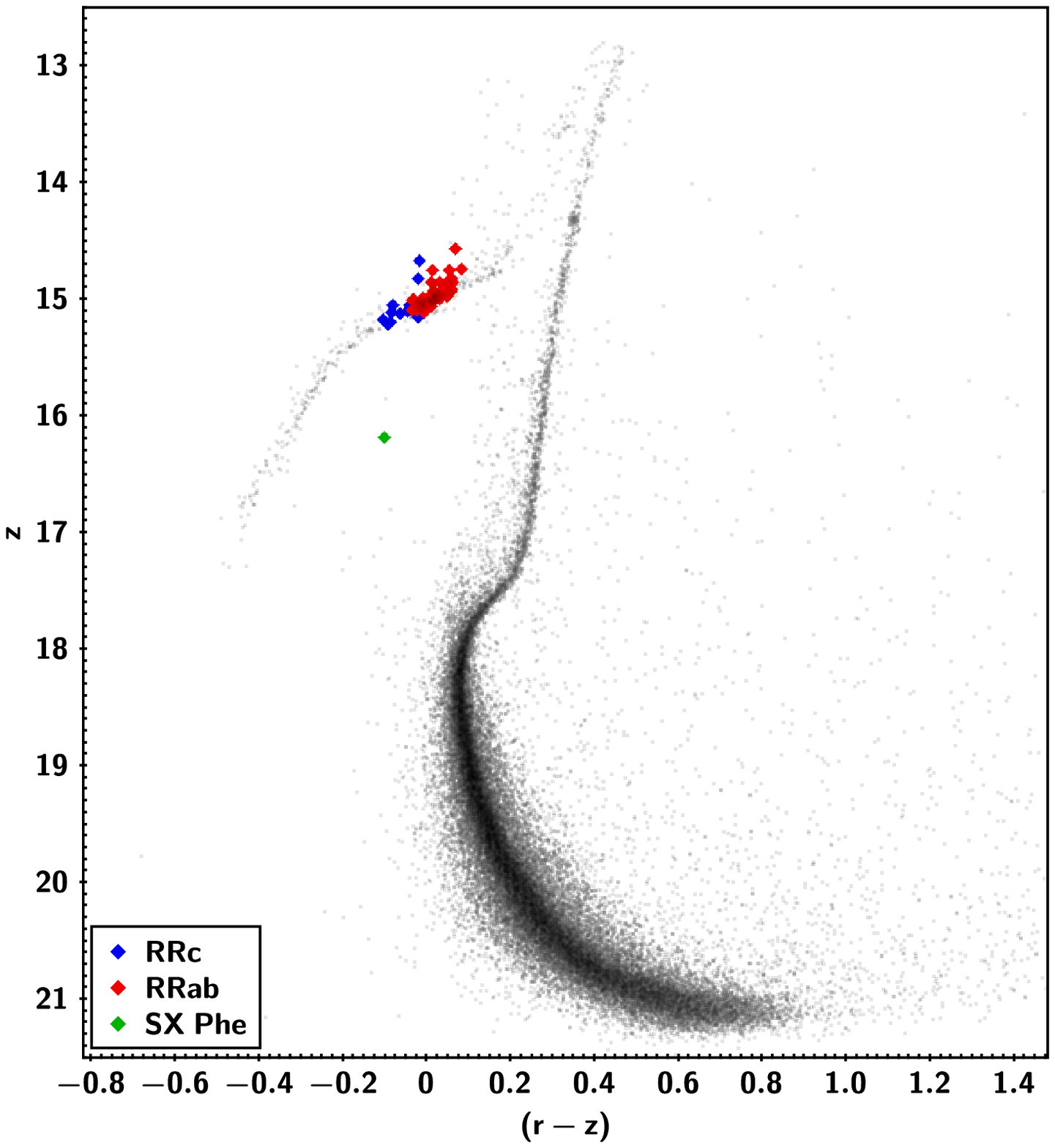}
\caption{Color-Magnitude diagrams ($g$ vs $(g-i)$, left; $z$ vs $(r-z)$, right) 
of stars inside a radius of $17\farcm 7$ ($10 \times r_h$) from the center of M5 (54,701 stars). 
Periodic variable stars are indicated with colored symbols.}
\label{fig-M5CMD}
\end{figure*}

Color-magnitude diagrams (CMD) of stars within $10\,r_h$ \citep[$r_h=1\farcm 77$,][]{harris96} are shown in
Figure~\ref{fig-M5CMD} using different filter combinations. These CMDs were built with the average magnitudes for each star in the catalog.
The large area covered by DECam results in very sharp features of the
cluster because of the high number of stars included in these diagrams. 
Notice however that,
due to the gap between CCDs and our constraint in photometry near the border of the CCDs, there is a strip of $1\farcm 38$ centered on the 
cluster that is not covered in our catalogs (see Figure~\ref{fig-M5sky}). The catalog associated with these diagrams can be obtained from 
Table~\ref{tab-M5catalog}. The $\sigma_\lambda$ in this table refers to the standard deviation of the mean of all the measurements for a star in each filter.

\floattable
\begin{deluxetable}{lcccccccccccc}
\tabletypesize{\small}
\tablecolumns{13}
\tablewidth{0pc}
\tablecaption{Photometry within a radius of $17\farcm 7$ from the center of M5 \label{tab-M5catalog}}
\tablehead{
ID &  RA (deg) & DEC (deg) & u & $\sigma_u$ & g & $\sigma_g$  & r & $\sigma_r$ & i & $\sigma_i$ & z & $\sigma_z$ \\
}
\startdata
 18709 & 229.46147 & 2.29986 & 16.438 &  0.006 & 15.264 &  0.005 & 14.692 &  0.005 & 14.532 &  0.006 & 14.452 &  0.003 \\
 18711 & 229.48143 & 2.28796 & 16.792 &  0.005 & 15.478 &  0.006 & 14.781 &  0.005 & 14.552 &  0.007 & 14.425 &  0.003 \\
 18712 & 229.48684 & 2.30266 & 16.082 &  0.005 & 14.999 &  0.004 & 14.531 &  0.005 & 14.431 &  0.008 & 14.395 &  0.004 \\
 18713 & 229.49202 & 2.28002 & 15.141 &  0.007 & 13.882 &  0.004 & 13.334 &  0.007 & 13.204 &  0.012 & 13.139 &  0.003 \\
 18714 & 229.60225 & 2.33137 & 16.374 &  0.011 & 15.347 &  0.015 & 14.874 &  0.011 & 14.768 &  0.006 & 14.732 &  0.010 \\
\enddata
\tablecomments{Table~\ref{tab-M5catalog} is published in its entirety in the electronic edition of The 
Astronomical Journal. A portion is shown here for guidance regarding its form and content.}
\end{deluxetable}  

The HB is quite horizontal in the $g$ band (Figure~\ref{fig-M5CMD}, left panel), 
while it has a significant slope in redder bands like the $z$ band shown in 
the right panel.
As expected, all RR Lyrae stars within the radius of $10\,r_h$ lie on top of the HB of the cluster. 
The obvious outlier below the HB (at $g\sim 16.2$)
is a SX Phe star with a period of only 0.0898d. 
Neither the SX Phe star nor the RR Lyrae stars outside that radius (Figure~\ref{fig-M5sky}) will be used 
in the following analysis.

\section{Minimum Light Colors of RR Lyrae Stars in M5}\label{sec-color}

In order to obtain the color at minimum light, we chose to measure the magnitude of the template fitted to each RR Lyrae star at phase $\phi=0.65$, as shown in 
Figure~\ref{fig-lightcurve}. Because not all of the light curves in 
our data are well sampled, the use of the template is particularly useful for cases where the lightcurve is lacking observations near minimum light and near maximum light, which is needed to define the initial phase ($\phi=0$). Each star was measured in all available bands
and the magnitudes at this phase are recorded in Table~\ref{tab-RRLS} as $m_{\rm ref}$. In the original work by \cite{Sturch66}, the minimum color is defined as the mean color between phases 0.5 and 0.80. We experimented with this definition as well but found no
significant difference with our values and decided to keep the single-phase definition. A similar conclusion was also reached by 
\citet{kunder10}.

We computed mean observed colors at minimum light of the 47 M5  {\rrab } in the DECam passband system and recorded the results in   
Table~\ref{tab-colors}. For comparison we also provide in Table ~\ref{tab-colors} the mean colors (the colors calculated with the mean magnitude in each band) and 
their standard deviations. 
The dispersion observed in the colors at minimum light are always lower than the ones from the mean colors, confirming that the former is
the one suitable to be used as a color standard.

As expected from the early works of \citet{Sturch66} in the $UBV$ bands and later works by \citet{mateo95,guldenschuh05} and \citet{kunder10} in $VRI$, the colors at minimum light of the fundamental mode RR 
Lyrae stars have only a very small dispersion.  The dispersion is particularly low in the infrared colors  as they 
are freer of line blanketing.  In particular, the $(r-i)_{\rm min}$ and $(i-z)_{\rm min}$ colors have dispersions of $\lesssim 0.02$ magnitudes, 
which are about the same
size as the observed photometric errors in the colors at the level of the HB of M5. 

Assuming a reddening of $E(B-V) = 0.035 \pm 0.005$ for M5 \citep{carretta00}, a standard reddening law of $R_V=3.1$ and the following extinction to reddening 
($A_\lambda/E(B-V)$) values\footnote{These coefficients slightly differ from those in \citet{schlafly11} since they are calculated using updated values for the effective DECam throughput (E. Schlafly, private communication)} for $ugriz$: 3.995, 3.214, 2.165, 1.592, 1.211 \citep{schlafly11}, 
we transformed those mean observed colors at minimum light to dereddened values (see Table~\ref{tab-colors}). 
The errors quoted in the 
dereddened colors at minimum light include both the observational errors at the level of the HB in each band and the error in the 
assumed extinction toward the cluster.

\floattable
\begin{deluxetable}{cccccc}
\tablecolumns{6}
\tablewidth{0pc}
\tablecaption{Mean and Minimum-Light colors of Fundamental-Mode RR Lyrae Stars in M5 \label{tab-colors}}
\tablehead{
Color & $(\langle X \rangle - \langle Y \rangle)$\tablenotemark{1}  &  Std Dev\tablenotemark{2}  &
$(X-Y)_{\rm min}$\tablenotemark{3}  &  Std Dev\tablenotemark{4}  & $(X_0 - Y_0)_{\rm min}$\tablenotemark{5}  \\
}
\startdata
$u-g$ & 0.575 & 0.059 & 0.644 & 0.034 & $0.616 \pm 0.031$ \\
$g-r$  & 0.157 & 0.055 & 0.288 & 0.035 & $0.251 \pm 0.025$  \\
$g-i$  & 0.147 & 0.076 & 0.331 & 0.041 & $0.274 \pm 0.024$ \\
$r-i$   & -0.010 & 0.023 & 0.043 & 0.017 & $0.022 \pm 0.021$ \\
$r-z$ & 0.016 & 0.029 & 0.082 & 0.023 & $0.049 \pm 0.020$ \\
$i-z$ & 0.026 & 0.011 & 0.039 & 0.016 & $0.026 \pm 0.020$ \\
\enddata
\tablenotetext{1}{Average of the observed mean colors}
\tablenotetext{2}{Standard deviation of the distribution of the observed mean colors}
\tablenotetext{3}{Average of the observed colors at minimum light}
\tablenotetext{4}{Standard deviation of the distribution of observed colors at minimum light}
\tablenotetext{5}{Average of the reddening-corrected colors at minimum light and its error}
\end{deluxetable}  

\subsection{Dependence of colors at minimum light with periods}

\begin{figure*}[htb!]
\plotone{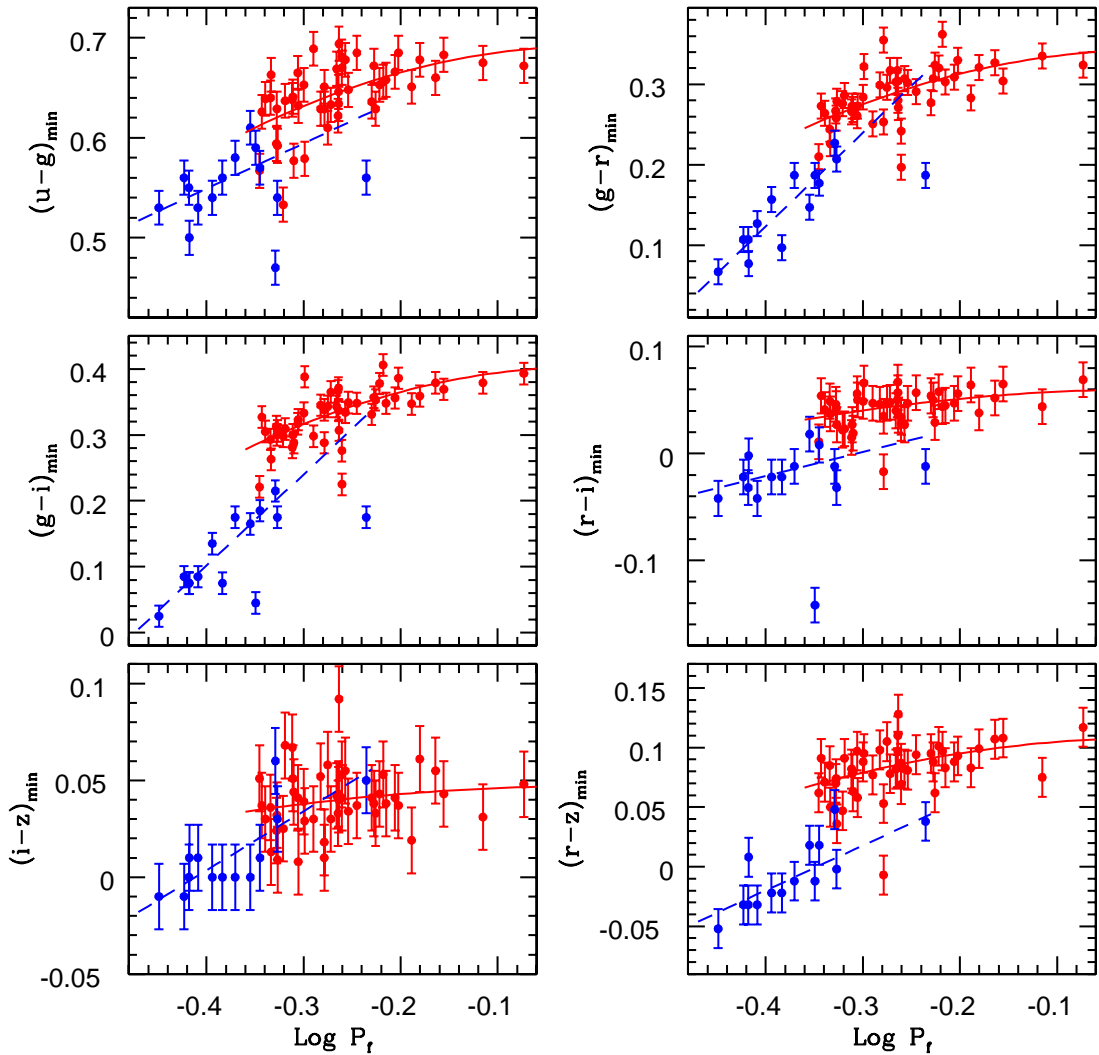}
\caption{Dependence of the colors at minimum light of the RR Lyrae stars in M5 with the logarithm of the fundamentalized period $P_f$.
Red symbols correspond to {\rrab } while blue symbols are the \rrc . 
The red solid line is a fit of the {\rrab } data, while the blue dashed line is a linear fit to the {\rrc } stars.}
\label{fig-periods}
\end{figure*}

A dependence on period may be the responsible for part of the dispersion observed in some of the colors at minimum light discussed above.
We checked for such dependence and present the results in Figure~\ref{fig-periods}, including this time the {\rrc } as well. Similar behavior 
as the one observed in Figure~\ref{fig-periods} is seen in the analysis of SDSS colors by \citet{ngeow17}.
Although for the colors from redder filters the dependence  on periods
seems to be small, for $(u-g)_{\rm min}, (g-r)_{\rm min}$, and
$(g-i)_{\rm min}$, it is large enough that appropriate corrections are
needed to obtain the true color at minimum light for an RR Lyrae star of a given period. We fit parabolic functions to the {\rrab }  data in each 
color $(X-Y)_{\rm min}$ as a function of the logarithm of the period (P) of the form:

\begin{equation}
(X-Y)_{\rm min} = a + b\, (\log P)^2
\end{equation}

The coefficients of such fits and their rms are shown in Table~\ref{tab-coefficients_ab}, and plotted as solid red lines in Figure~\ref{fig-periods}.
The standard deviation values and the rms of the fits in Tables~\ref{tab-colors} and \ref{tab-coefficients_ab} suggests that better results can 
be obtained when using colors from redder DECam filters to estimate interstellar extinction toward RR Lyrae stars as the rms of the fitted
functions are the smallest in these bands.  When taking into account the period,
the use of period-corrected minimum colors may lead to errors in the true color of  $\lesssim 0.016$ mags in $(r-i)_{\rm min}$, $(r-z)_{\rm min}$, and $(i-z)_{\rm min}$. The coefficients in Table~\ref{tab-coefficients_ab} can then be used to obtain the intrinsic colors of {\rrab } of
a given period and, thus, to determine accurate line of sight reddening toward those stars.

Two {\rrab } were left out from these fits because they showed discrepant behaviors (see Fig~\ref{fig-periods}): 74799 and 74542. Our data for the former is
consistent with a sinusoidal light curve and small amplitude. The
colors for this star are too red to be a \rrc. \citet{arellano16} flagged this
star (V85 in their nomenclature) as a Blazhko star, but their lightcurve looks clearly like an \rrab. Thus, it may be a case of extreme Blazhko effect or mode changing. On the other hand, 
star 74542 has very few points in the $g$ and $r$ lightcurves ($<10$) and that may have produced discrepant colors.

In the past, \citet{kanbur05} and \citet{kunder08} reached the conclusion that no strong 
dependence on period existed for RR Lyrae stars measured in the ($V-R$) color. This is still 
almost true (very small dependence) in colors combining the $riz$ DECam filters, 
but the dependence is important in colors that include the $u$ or $g$ filters.
We note that our photometry is more precise than the MACHO photometry used in \citet{kanbur05}, 
and our sample size is larger (more than twice as large) as that in \citet{kunder08}, 
which likely contributes to the reason we are able to see this small trend.

\begin{deluxetable}{ccccc}
\tablecolumns{5}
\tablewidth{0pc}
\tablecaption{Coefficients and rms of the parabolic fits to the Log (P) - Color at Minimum Light relationship for \rrab \label{tab-coefficients_ab}}
\tablehead{
Color & a & b & rms & E(X-Y)\tablenotemark{1} \\
}
\startdata
$u-g$ & 0.692 & -0.669 & 0.029 & 0.027 \\
$g-r$ & 0.343 & -0.753 & 0.022 & 0.037 \\
$g-i$ & 0.404 & -0.973 & 0.026 & 0.057 \\
$r-i$ & 0.060 & -0.220  & 0.012 & 0.020 \\
$i-z$ & 0.047 & -0.102 & 0.016 & 0.033 \\
$r-z$ & 0.108 & -0.322 & 0.016 & 0.013 \\
\enddata
\tablenotetext{1}{Excess color corresponding to $E(B-V)=0.035$ \citep{carretta00} using the extinction to reddening coefficients given in \S\ref{sec-color}}
\end{deluxetable}  

We also explored the behavior of the color at minimum light of {\rrc } in M5, measured at the phase corresponding to the lowest point of their light curves. 
For these stars we used the fundamentalized period, $P_f$, as defined in \citet{catelan09}:

\begin{equation}
\log P_f = \log P + 0.128
\label{eq-pf}
\end{equation}

The {\rrc } are plotted with blue symbols in Figure~\ref{fig-periods}. Given the observed behavior of these stars in the plots and because there are only 14 
{\rrc } in our data, we decided to fit only linear relationships. In
the figure it can be seen that there are some discrepant points in
most of the panels that were left out from the fit. They corresponded to stars 74561, 72475, and 74395.  {\rrc } stars are hard to separate unambiguously from contact binaries and
contamination by those stars may be causing some of the observed discrepancies. The results of the linear fits are 
shown as blue dashed lines in Figure~\ref{fig-periods} and the slope, intercept, and rms are reported in Table~\ref{tab-coefficients_c}.

From Figure~\ref{fig-periods} it is clear that {\rrc } do not follow well the trend observed for the \rrab. A small offset in the colors is 
observed between the two types of RR Lyrae stars and, thus, it is more
convenient to use separate relationships for each type. As in the case
of \rrab, the smallest dispersions in the Color-$\log P_f$ relationships for {\rrc } were also obtained when using colors constructed with the redder filters $r$, $i$, and $z$.

Notice that the coefficients given in Table~\ref{tab-coefficients_ab} and ~\ref{tab-coefficients_c} provide the colors at 
minimum light of the RR Lyrae stars in M5. If they are to be used for stars in any other part of the sky, they should be corrected for reddening first. For convenience, we provide, in the
last column of Table~\ref{tab-coefficients_ab}, the 
excess colors for the different DECam colors assuming an $E(B-V) =0.035$ for M5 \citep{carretta00} and the reddening law and coefficients given above.

\begin{deluxetable}{cccccc}
\tablecolumns{4}
\tablewidth{0pc}
\tablecaption{Coefficients and rms of the Linear Fits to the $\log (P_f)$ - Color at Minimum Light relationship for \rrc  \label{tab-coefficients_c}}
\tablehead{
Color & Intercept & Slope & rms \\
}
\startdata
$u-g$ & 0.730 & 0.453 & 0.026 \\
$g-r$ &  0.589 & 1.165 & 0.023 \\
$g-i$ & 0.654 & 1.381 & 0.023 \\ 
$r-i$ & 0.069 & 0.226 & 0.017 \\
$i-z$ & 0.126 & 0.306 & 0.014 \\
$r-z$ & 0.131 & 0.376 & 0.015 \\
\enddata
\end{deluxetable}  

\subsection{Dependence on Metallicity}

We need also to understand the dependence of the color at minimum
light on metallicity.  \citet{Sturch66} reported a non-negligible
dependence of $B-V$ with metallicity, with $U-B$ showing a much larger
dependence. He was thus able to mitigate the effect on $B-V$ by
referring to  $U-B$.  This suggests that the metallicity dependence
arises either wholly, or in most part, due to line blanketing. If so,
by using redder colors, where the line blanketing is much smaller, we
expect to further mitigate the effect.  To get a quantitative
estimate, compare Table~8 in \citet{jones87a} with Table~4 in
\citet{jones87b}. In the former, synthetic colors are calculated for
various temperatures and gravities for [Fe/H]$= -2.2$ and in the
latter, the same for [Fe/H]$= -0.75$. We see that while the predicted
$B-V$ colors at the same temperature and gravity differ by as much as
0.06 mag for temperatures in the range 5500K and 6500K, while corresponding $V-K$ colors differ by only 0.02 mag.   Further, Figure~2 in \cite{blanco92b} (where he re-examines the systematics and accuracy of Sturch's original relations) indicates that the metallicity corrections for $B-V$ are independent of the temperature, in the range 5500K to 6500K.  

\begin{figure}[htb!]
\plotone{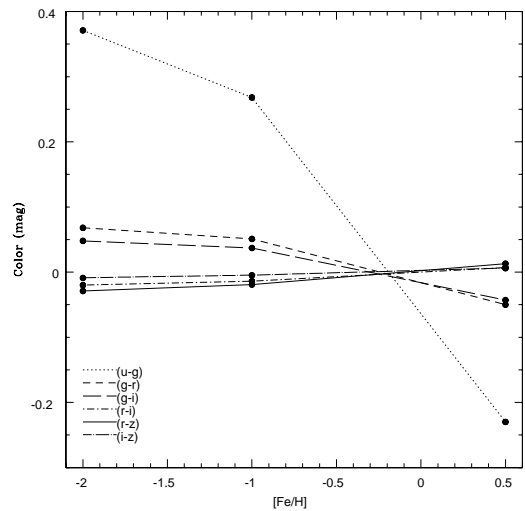}
\caption{Dependence of different DECam colors for a synthetic star ($T_{\rm eff} = 6250$ K, $\log g = 2.0$) with metallicity.}
\label{fig-metallicity}
\end{figure}

With this preface, to explore the metallicity dependence of colors from DECam at minimum light, we used the flux from a  Kurucz model\footnote{as presented in ftp://ftp.stsci.edu/cdbs/grid/k93models} with  $T_{\rm eff} = 6250$ K, $\log g = 2.0$, and three different metallicities: 
[Fe/H]$=-1.0, -2.0, +0.5$. We convolved these spectra with the DECam filter throughput 
curves\footnote{\url{http://www.ctio.noao.edu/noao/content/Dark-Energy-Camera-DECam}} 
and calculated the color of the synthetic RR  Lyrae star at each
metallicity (Figure~\ref{fig-metallicity}). The chosen temperature and gravity are representative of a fundamental RR Lyrae star near minimum light (in unpublished work, A. Saha has ascertained that this choice is consistent with both the SED \emph{including the Balmer jump}, as well as the Balmer line profiles for 6 bright RR Lyrae stars spanning 2 dex in metallicity: U~Lep, RX~Eri, ST~Oph, U~Pic, SV~Eri and HH~Pup while in their low-light phases). In any case, as Figure~2 in \citet{blanco92} indicates, the exact choice of temperature should not significantly alter the derived dependencies. 
 
The results of these calculations indicate that the colors $(r-i)$, $(r-z)$, and $(i-z)$ may have maximum variations of 0.027, 0.042 and 0.015 mags, respectively, 
in the full metallicity range explored. 
These differences are even smaller in the range of metallicity from $-2 < {\rm [Fe/H]} < -1$, where
the differences in color amount to only 0.006, 0.010 and 0.004 in $(r-i)$, $(r-z)$, and $(i-z)$, respectively.

The variations of the $(u-g)$ color with metallicity are large enough
to make it not suitable for use as a standard color measurement if the metallicity of the
RR Lyrae star is not known. These large variations have been observed before, in the $UBV$ system by \citet{Sturch66} and others. This is expected, because metallicity expresses itself through line-blanketing and opacity effects in the blue ($u$ and $g$) and ultra-violet, but much less so as one goes progressively to the red.   Colors involving the $g$ filter ($g-r$ and $g-i$) have variations of colors in the range of metallicity studied as large as  0.1 magnitudes. 

\section{Absolute Magnitudes of M5 RR Lyrae Stars in DECam filters}\label{sec-M}

\begin{figure*}[htb!]
\plotone{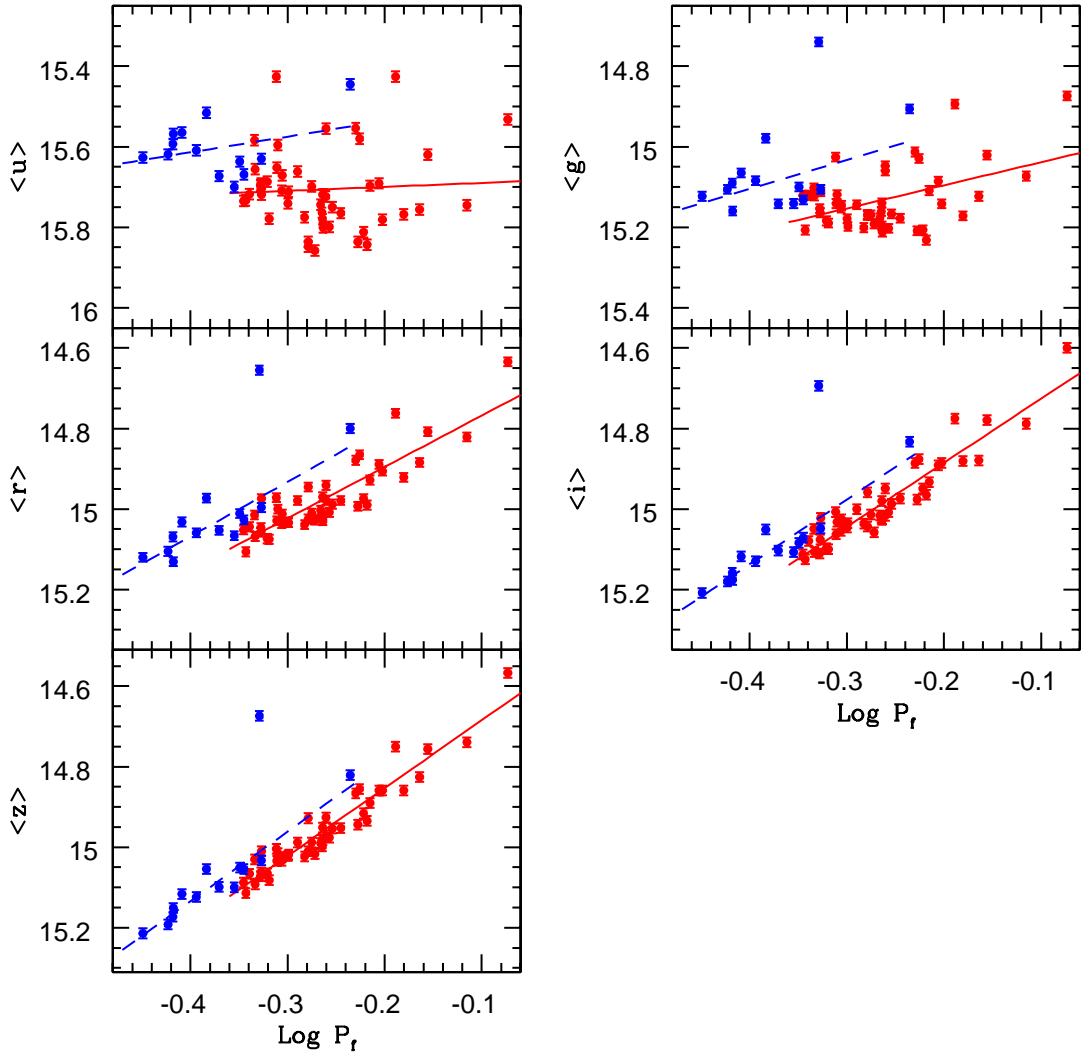}
\caption{Dependence of the mean magnitude in each DECam band as a function of period for the {\rrab } (red symbols) and
{\rrc } (blue symbols) in M5.  The solid red line is a least square fit for the {\rrab } stars while the dashed blue line is the best fit for the {\rrc } stars.}
\label{fig-absmag}
\end{figure*}

Calibration of the absolute magnitudes of RR Lyrae stars in the DECam filters does not exist to date.
For SDSS passbands, Period-Luminosity (P-L) relationships have been
provided by \citet{caceres08} that are based on theoretical models.
Here we determine the empirical values of the absolute magnitudes of the RR Lyrae stars in M5 in the DECam natural filter system in which our data
have been calibrated. In Figure~\ref{fig-absmag} we show the observed mean magnitude of the RR Lyrae stars as a function of the logarithm
of their pulsation period. As before, the period for the {\rrc } stars has been "fundamentalized" using Equation~\ref{eq-pf}.
For all the filters, the {\rrc } stars seem to be slightly brighter and thus we fit P-L relationships separately for each type.
The best linear fits are given in Table~\ref{tab-PL}. 
The dependence of absolute magnitude with period in $r, i,$ and $z$ is strong but very well constrained, as implied by the low rms values of the linear fits.
In the $u$ and $g$ bands the correlation between mean magnitude and period is less obvious and the resulting fit has larger residuals. In any case, the slopes 
of the relationships  in these filters seem to be small, which is a consequence of the HB being mostly horizontal in these bands 
(see for example the CMD in Figure~\ref{fig-M5CMD}).
The {\rrab } stars have tighter relationships than the {\rrc } stars.

\begin{deluxetable}{ccccccc}
\tablecolumns{7}
\tablewidth{0pc}
\tablecaption{Coefficients and rms of the Linear Fits to the mean magnitude - $\log (P_f)$ relationship for RR Lyrae stars in M5 \label{tab-PL}}
\tablehead{
Band & \multicolumn{3}{c}{\rrab} & \multicolumn{3}{c}{\rrc} \\ 
         & Intercept & Slope & rms & Intercept & Slope & rms \\
}
\startdata
u & 15.68 & -0.10 & 0.07 & 15.46 & -0.39 & 0.13\\
g & 14.98 & -0.57 & 0.04 & 14.82 & -0.72 & 0.12 \\
r & 14.64 & -1.28 & 0.03 & 14.53 & -1.35 & 0.08 \\
i & 14.57 & -1.59 & 0.02 & 14.49 & -1.61 & 0.06 \\
z & 14.52 & -1.68 & 0.02 & 14.44 & -1.73 & 0.05 \\
\enddata
\end{deluxetable}  

\citet{layden05} determined the true distance modulus of M5 using main-sequence fitting and obtained $(m - M)_0 = 14.45 \pm 0.11$ mag, a value that is consistent with $14.44 \pm 0.02$ given by \citet{coppola10} based on IR observations of RR Lyrae stars. 
Assuming the latter and an excess color $E(B-V) =0.035\pm0.005$ \citep{carretta00}, we scaled the intercept parameters in Table~\ref{tab-PL} to obtain the 
following absolute P-L relationships. The error in the distance modulus and interstellar extinction have been added in quadrature to the intercept term.

\begin{eqnarray}
M_u \; (ab) &= (-0.10 \pm 0.24) \, \log P  + (1.10 \pm 0.13) \nonumber  \\
M_u \; (c) &= (-0.38 \pm 0.36) \, \log P_f  + (0.88 \pm 0.18)  \nonumber \\
M_g \; (ab) &= (-0.57 \pm 0.17) \, \log P  + (0.43 \pm 0.12) \nonumber \\
M_g \; (c) &= (-0.72 \pm 0.32) \, \log P_f  + (0.27 \pm 0.16)  \nonumber \\
M_r \; (ab) &= (-1.28 \pm 0.11) \, \log P  + (0.12 \pm 0.11)  \\
M_r \; (c) &= (-1.35 \pm 0.21) \, \log P_f  + (0.01 \pm 0.13)  \nonumber \\
M_i \; (ab) &= (-1.59 \pm 0.09) \, \log P  + (0.07 \pm 0.11) \nonumber  \\
M_i \; (c) &= (-1.61 \pm 0.16) \, \log P_f  + (0.00 \pm 0.12)  \nonumber \\
M_z \; (ab) &= (-1.68 \pm 0.08) \, \log P  + (0.03 \pm 0.11)  \nonumber \\
M_z \; (c) &= (-1.73 \pm 0.14) \, \log P_f  - (0.04 \pm 0.12) \nonumber 
\end{eqnarray}

Notice that although it is expected that absolute magnitudes also have a dependence on the metallicity, we are dealing here with RR Lyrae stars within a single 
globular cluster with a very small metallicity dispersion. The metallicity of M5 is [Fe/H]$=-1.25\pm0.05$ \citep{dias16}. 

For comparison, we calculated the predicted values of the absolute
magnitude in $i$ and $z$ for a RR Lyrae star with $P=0.5$d, using the
relationships given for the SDSS system by \citet{caceres08} that are based on synthetic spectra of RR Lyrae stars. To do this, we assumed [Fe/H]$=-1.25$ and [$\alpha$/Fe]$=0.24$ \citep{dias16}. Once the resulting values are transformed to the DECam system, the discrepancy between empirical absolute magnitudes and the ones derived from the equations in \citet{caceres08} are only 0.02 and 0.03 mag in $i$ and $z$ respectively. 
We notice that \citet{caceres08} were unable to produce relationships for $M_u$, $M_g$, and $M_r$ because they were not tight enough.

\section{Conclusions}\label{sec-conclu}

In this paper we provide colors at minimum light and absolute magnitudes of RR Lyrae stars (both {\rrab } and {\rrc }) in the $ugriz$ DECam system based on
observations of the Galactic globular cluster M5. The colors at minimum light of RR Lyrae stars constitute an important tool to determine the reddening along the line 
of sight up to the distance of the star.  This is our main motivation for the present study because these calibrations will be applied to a large scale survey
of RR Lyrae stars in the Bulge in forthcoming papers.

We studied the behavior of different DECam colors at minimum light with period and conclude that the reddest colors, $(r-i)_{\rm min}$, 
$(r-z)_{\rm min}$, and $(i-z)_{\rm min}$, are the best
options to be used as color standards. 
The dependence of color at minimum light with period is well behaved 
and we were able to fit functions to both {\rrab } and {\rrc } stars with resulting rms values $<0.02$ magnitudes. 

A caveat of this work however is the use of a single cluster with a fixed metallicity ([Fe/H]$=-1.25$). We tested to some extent the possible 
dependence on metallicity by analyzing synthetic spectra in the range 
$-2.0 <$ [Fe/H] $<+0.5$ and conclude that the color variations are at most 0.02, 0.03, and 0.01 magnitudes in that range of metallicity for
$(r-i)_{\rm min}$, $(r-z)_{\rm min}$, and $(i-z)_{\rm min}$, respectively. Also, from previous works based on $VRI$ magnitudes 
\citep{guldenschuh05,kunder10} we do not expect a strong dependence of colors at minimum light with metallicity. 
It will be desirable, however, to study other globular clusters with a range of metal content in order to confirm this statement. On the other hand, our main motivation for this work is to apply these calibrations to
RR Lyrae stars in the Galactic Bulge, which is in a metallicity regime not very different to that of M5 \citep[for example,][]{walker90}.

We also provide an empirical calibration of the absolute magnitude of the M5 RR Lyrae stars in the DECam system. Again, the reddest filters, $riz$ are the
best options for obtaining a good absolute magnitude (and hence, distance) since they produce tight P-L relationships for the {\rrab } stars,
with dispersion amounting to 0.03, 0.02, and 0.02 magnitudes in $r$, $i$, and $z$, respectively. The dispersions obtained for the {\rrc } in those bands are
also small, $<0.1$ magnitudes.

With its large FOV, DECam is an excellent instrument to study RR Lyrae stars over large extensions in the sky. We anticipate the calibrations provided in this paper 
will be useful for different projects that use this type of variables
as standards for distance and color.

\begin{acknowledgments} 

Based on observations at Cerro Tololo Inter-American Observatory, National Optical Astronomy Observatory (NOAO Prop. ID: 2013A-0719; PI: Saha), which is operated by the Association of Universities for Research in Astronomy (AURA) under a cooperative agreement with the National Science Foundation. 
This project used data obtained with the Dark Energy Camera (DECam), which was constructed by the Dark Energy Survey (DES) collaboration.
Funding for the DES Projects has been provided by 
the U.S. Department of Energy, 
the U.S. National Science Foundation, 
the Ministry of Science and Education of Spain, 
the Science and Technology Facilities Council of the United Kingdom, 
the Higher Education Funding Council for England, 
the National Center for Supercomputing Applications at the University of Illinois at Urbana-Champaign, 
the Kavli Institute of Cosmological Physics at the University of Chicago, 
the Center for Cosmology and Astro-Particle Physics at the Ohio State University, 
the Mitchell Institute for Fundamental Physics and Astronomy at Texas A\&M University, 
Financiadora de Estudos e Projetos, Funda{\c c}{\~a}o Carlos Chagas Filho de Amparo {\`a} Pesquisa do Estado do Rio de Janeiro, 
Conselho Nacional de Desenvolvimento Cient{\'i}fico e Tecnol{\'o}gico and the Minist{\'e}rio da Ci{\^e}ncia, Tecnologia e Inovac{\~a}o, 
the Deutsche Forschungsgemeinschaft, 
and the Collaborating Institutions in the Dark Energy Survey. 
The Collaborating Institutions are 
Argonne National Laboratory, 
the University of California at Santa Cruz, 
the University of Cambridge, 
Centro de Investigaciones En{\'e}rgeticas, Medioambientales y Tecnol{\'o}gicas-Madrid, 
the University of Chicago, 
University College London, 
the DES-Brazil Consortium, 
the University of Edinburgh, 
the Eidgen{\"o}ssische Technische Hoch\-schule (ETH) Z{\"u}rich, 
Fermi National Accelerator Laboratory, 
the University of Illinois at Urbana-Champaign, 
the Institut de Ci{\`e}ncies de l'Espai (IEEC/CSIC), 
the Institut de F{\'i}sica d'Altes Energies, 
Lawrence Berkeley National Laboratory, 
the Ludwig-Maximilians Universit{\"a}t M{\"u}nchen and the associated Excellence Cluster Universe, 
the University of Michigan, 
{the} National Optical Astronomy Observatory, 
the University of Nottingham, 
the Ohio State University, 
the University of Pennsylvania, 
the University of Portsmouth, 
SLAC National Accelerator Laboratory, 
Stanford University, 
the University of Sussex, 
and Texas A\&M University.
EO was partially supported by the NSF through grant AST-1313006.

\software{Community Pipeline \citep{Valdes14}, DoPHOT \citep{Schechter93}, Psearch \citep{saha17}, ATLAS (Kurucz 1993), TOPCAT, IDL, MySQL}

\facility{Blanco (DECam)}

\end{acknowledgments}



\end{document}